# Ultrafast dynamics of vibrational symmetry breaking in a charge-ordered nickelate

G. Coslovich,[1,2*] A.F. Kemper,[3,4] S. Behl,[1] B. Huber,[1] H. A. Bechtel,[5] T. Sasagawa,[6] M. C. Martin,[5] A. Lanzara,[1,7] and R. A. Kaindl[1*]

[1]*Materials Sciences Division, E. O. Lawrence Berkeley National Laboratory, 1 Cyclotron Road, Berkeley, CA 94720, USA.*
[2]*SLAC National Accelerator Laboratory, 2575 Sand Hill Road, Menlo Park, CA 94025, USA.*
[3]*Computational Research Division, E. O. Lawrence Berkeley National Laboratory, 1 Cyclotron Road, Berkeley, CA 94720, USA.*
[4]*Department of Physics, North Carolina State University, Raleigh, NC 27695, USA.*
[5]*Advanced Light Source, E. O. Lawrence Berkeley National Laboratory, 1 Cyclotron Road, Berkeley, CA 94720, USA.*
[6]*Materials and Structures Laboratory, Tokyo Institute of Technology, Kanagawa 226-8503, Japan*
[7]*Department of Physics, University of California, Berkeley, CA 94720, USA.*
*\*Corresponding author. E-mail: gcoslovich@lbl.gov (G.C.), rakaindl@lbl.gov (R.A.K.)*

**The ability to probe symmetry breaking transitions on their natural time scales is one of the key challenges in nonequilibrium physics. Stripe ordering represents an intriguing type of broken symmetry, where complex interactions result in atomic-scale lines of charge and spin density. Although phonon anomalies and periodic distortions attest the importance of electron-phonon coupling in the formation of stripe phases, a direct time-domain view of vibrational symmetry breaking is lacking. We report experiments that track the transient multi-THz response of the model stripe compound $La_{1.75}Sr_{0.25}NiO_4$, yielding novel insight into its electronic and structural dynamics following an ultrafast optical quench. We find that although electronic carriers are immediately delocalized, the crystal symmetry remains initially frozen – as witnessed by time-delayed suppression of zone-folded Ni-O bending modes acting as a fingerprint of lattice symmetry. Longitudinal and transverse vibrations react with different speeds, indicating a strong directionality and an important role of polar interactions.**

**The hidden complexity of electronic and structural coupling during stripe melting and formation, captured here within a single terahertz spectrum, opens new paths to understanding symmetry breaking dynamics in solids.**

**INTRODUCTION**

The dynamic coordination of atomic structure and electronic states plays a pivotal role in many phenomena of nature, helping determine the energetic pathways of chemical reactions, the conformational function of biomolecules, or the nature of symmetry-broken phases in solids (*1-3*). In particular, complex quantum materials are characterized by the correlated interplay of electronic and vibrational excitations that spawns a wealth of emerging phases. However, the conventional slow adiabatic tuning of phase transitions masks the causal time ordering of the intrinsic interactions that evolve on short timescales. This has motivated time-resolved studies of phase transitions, uncovering remarkably fast cooperative electronic and structural dynamics (*4-7*).

In particular, complex oxides exhibit an intriguing self-organization of atomic-scale charge, spin and vibrational patterns which can fluctuate in time and space (*8-11*). Recent observations of stripe order in cuprates attracted much attention (*12-14*), and vibrational excitation of stripe-ordered phases revealed surprising signatures of light-induced superconductivity (*15*). Moreover, the role of electron-phonon coupling in these unconventional charge density waves remains under discussion (*16, 17*). Understanding the dynamic forces motivates ultrafast studies to elucidate how stripes melt, form, and fluctuate. The non-superconducting nickelate family $La_{2-x}Sr_xNiO_4$ (LSNO) represents an ideal model system for isolating stripe dynamics, exhibiting distinct lattice and spin ordering but only weak screening of low-energy excitations (*11, 18*). Recent ultrafast studies demonstrated the ability to optically suppress stripes in nickelates, observing rapid charge localization and multi-picosecond formation of its long-range electronic order (*19-21*). However, the dynamic coupling between the periodic electronic and vibrational modulations during melting and symmetry-breaking formation of stripes has remained elusive so far.

**RESULTS**

To gain direct insight into the degree of symmetry breaking on fundamental time scales, we track the ultrafast terahertz reflectivity dynamics of the Ni-O bending mode in LSNO, which exhibits features arising from the stripe-phase order. Figure 1A shows the equilibrium in-plane optical conductivity spectra $\sigma_1(\omega)$ around the bending-mode phonon (≈11 THz) of our $La_{1.75}Sr_{0.25}NiO_4$ single crystal as obtained from broadband Fourier-transform infrared (FTIR) spectroscopy. Above the charge-ordering transition temperature $T_{CO} \approx 105$ K, this vibration is described by a single oscillator on top of a flat and featureless background. The background arises from itinerant charge transport, resulting in the broad electronic conductivity that progressively vanishes as the crystal is cooled below ≈250 K because of the onset of short-range charge localization (*21*). Notably, as shown in Fig. 1A additional sharp resonances appear as the temperature drops into the stripe phase (*22, 23*). We have determined the oscillator energies and strengths from a multi-oscillator fit (lines in Fig. 1A), with the corresponding energies plotted in Fig. 1B. In the charge-ordered state, the response comprises three sharp peaks at energies $\omega_1$ to $\omega_3$, on top of a remnant of the broad high-temperature mode ($\omega_0$) whose frequency continues smoothly across the whole temperature range. As shown in Fig. 1C, the strength of the original phonon line at $\omega_0$ is suppressed as the new vibrational degrees-of-freedom emerge (indicated here for $\omega_2$, for all modes see fig. S5), underscoring a redistribution of spectral weight and connection to the stripe-phase order.

The appearance of new resonances in the vibrational spectrum can be understood as arising from momentum back-folding of the phonon dispersion (*23*). Figure 1D shows the $E_u$ bending mode dispersion of the $La_2NiO_4$ lattice along the relevant directions in momentum space, as calculated from DFT combined with the finite displacement method (see the Materials and Methods). The dispersion is split into a higher-energy LO and a lower-energy TO branch. In the presence of a superlattice modulation the Brillouin zone size is reduced, which folds vibrations at the charge-order wavevector $q_{CO} = (0.55,0,1)$ back to the zone center at Γ resulting in new IR-active modes. On the basis of the dispersion we identify the highest resonance ($\omega_1$) with the LO vibration back-folded from $q_{CO}$ and the sharp intermediate resonance ($\omega_2$) as the corresponding zone-folded TO mode (see Fig. 1D). In turn, we assign the weak $\omega_3$ peak to the TO mode at $2q_{CO}$, whereas the corresponding second-order LO mode is masked by the strong $\omega_1$ peak. Because the LSNO crystal exhibits an intrinsic distribution of 90°-rotated nanoscale stripe domains across the sample (*11*), an incident terahertz wave in our geometry is variably oriented either parallel or perpendicular to each domain's stripes (see the Materials and Methods). Accordingly, it can sense both the zone-folded TO vibrations, which are polarized transverse to $q_{CO}$ and thus parallel to the stripes, and the corresponding LO mode polarized perpendicular to the stripes. We note that although the LO mode at $q_{CO}$ is microscopically of longitudinal character, zone-folding back to q≈0 results in an average polarization in each domain to which the transverse terahertz wave couples. The sharp resonances are thus a direct consequence of stripe-phase symmetry breaking, yielding a sensitive vibrational fingerprint of the periodic lattice distortion, in contrast to resonant x-ray diffraction (XRD) (*19*) at the Ni L-edge which scatters off electronic density modulations..

To reveal the dynamical forces acting between the electronic and vibrational constituents of the stripe-phase modulation, we quench the electronic order with a femtosecond near-IR pulse exciting interband electronic transitions (*21*) and resonantly probe the evolving terahertz reflectivity changes around the Ni-O bending vibration. In these experiments, phase-matched terahertz generation and electro-optic sampling in GaSe are used to optimally center and overlap the spectrum of the multi-terahertz probe with the bending mode resonance (Fig. 2A). The reference terahertz field $E(t)$ reflected off the LSNO crystal at $T = 30$ K is shown as the black line in Fig. 2B. Our experimental scheme allows for scanning the femtosecond probe fields within a large time window of up to 8 ps, corresponding to a high spectral resolution $\Delta\nu < 0.2$ THz that is necessary to resolve the sharp multiplet of bending mode resonances. The resulting photoinduced changes $\Delta E(t)$ of the terahertz probe are shown in Fig 2B as blue lines. To retain a high time resolution, the transient terahertz traces are sampled at fixed pump-probe delays $\Delta t$ between near-IR excitation and electro-optic detection using the scheme of Kindt and Schmuttenmaer (*24*) (see the Materials and Methods and section S2). With increasing delay $\Delta t$, the terahertz traces develop a complex ringing shape at later times, indicating the presence of spectrally narrow features in the LSNO response.

The terahertz spectral response is obtained from the amplitude and phase of the Fourier-transformed field traces, taking into account the reflective sample geometry and excitation conditions (see the Materials and Methods). Figure 3A maps out the resulting time evolution of the LSNO optical conductivity changes. Directly after excitation we observe a broadband conductivity increase, whereas spectrally sharp features emerge at later pump-probe delays. These sharp modulations are located around the energies of the three stripe-phase phonon peaks. The transient response corresponds to a suppression of these zone-folded Ni-O resonances, as evident in Fig. 3B from the similarity in frequency and sign of the oscillations in the transient spectral change compared to the difference expected from thermal stripe melting. Therefore the multi-terahertz probe directly evidences the loss of the stripe-phase vibrational superlattice on an ultrafast time scale.

For quantitative insight, we consider the corresponding spectral cuts of the pump-induced conductivity $\Delta\sigma_1$ and dielectric function change $\Delta\varepsilon_1$, which are shown in Fig. 3 (C and D) for several pump-probe delays. These spectra can be analyzed by differential changes of the multi-oscillator model, used above to describe the equilibrium response, whose parameters are constrained by the need to fit both real and imaginary parts of the conductivity change $\Delta\sigma(\omega) = \Delta\sigma_1(\omega) + i\omega\varepsilon_0 \cdot \Delta\varepsilon_1(\omega)$. The resulting model curves – lines in Fig. 3 (C and D) – closely describe the experimental data. Here, the broadband conductivity rise directly after excitation ($\Delta t = 0$ ps) represents a fast electronic response that decays completely within $\approx 1$ ps. With increasing time delay, the spectra become dominated by the appearance of spectrally sharp modulations. As the model confirms, these narrowband conductivity changes are reproduced by a concomitant blueshift and strong reduction of the zone-folded phonon resonances.

Figure 4 (A and B) shows the time dependence of the electronic spectral weight and the peak strengths of the zone-folded phonons (filled dots), as obtained from the multi-oscillator fits. The electronic response appears immediately with photoexcitation followed by a $\approx 0.4$ ps

single-exponential decay. This fast dynamics and the related broadband spectral response ($\Delta t = 0$ ps, Fig. 3C) arise from rapid delocalization and localization of itinerant hole carriers, in agreement with previous studies of the mid-IR pseudogap dynamics in LSNO with the fluence taken into account (see fig. S6) (*21*). In stark contrast, the dynamics of the zone-folded Ni-O vibrations is significantly slower, as evident from Fig. 4B. The LO superlattice peak exhibits a sub-picosecond time-delayed suppression, resulting in the reduction of its peak strength by ≈25%. Notably, the zone-folded TO vibration evolves even slower (Fig. 4B) and takes almost 2 ps to reach its fullest peak suppression, with analogous slow dynamics in the higher-order transverse mode (see fig. S7). For all modes, these vibrational changes revert on a multi-picosecond timescale, because the crystal symmetry breaking reestablishes alongside the long-range stripe formation.

**DISCUSSION**

Thus, the transient terahertz response reveals nontrivial dynamics of the stripe-phase vibrational order in LSNO under nonequilibrium conditions. Ultrafast excitation increases the excess energy of the electronic carriers, resulting in a quasi-instantaneous charge delocalization and ensuing partial melting of the electronic stripes as reported previously (*19-21*). At early times ($\Delta t = 0$ ps) the crystal distortions still remain "frozen" in their stripe-phase symmetry, whereas the charges have become itinerant. Within a few 100 fs the charges localize, which contributes to disrupting the electronic order and may further involve dissipation into strongly-coupled phonons analogous to other complex oxides (*25, 26*). Here, the longitudinal zone-folded distortions adapt on a similar time scale, whereas the transverse distortions remain locked in a symmetry-broken state. This indicates a situation where the translational symmetry is restored along the ordering wave vector for only one of the two modes. The disparity is also directly evident in the transient terahertz spectra (Fig. 3A), underscoring the different time evolution of the LO superlattice signature around 380 cm$^{-1}$ and that of the TO-related superlattice vibrations around 335 and 360 cm$^{-1}$. This physical behavior deviates significantly from a thermal scenario and represents instead a distinct hallmark of the nonequilibrium melting of stripe order in LSNO.

A further surprising aspect of the nonthermal phase transition is found in the transient blueshift of all the folded phonons, as shown in Fig. 4C. The shift closely matches the dynamics of the corresponding peak strengths. This observation contrasts with the overall equilibrium trend for the Ni-O bulk bending vibration ($\omega_0$ mode; line in Fig. 1B) which redshifts with rising temperature. We suggest that this effect originates from a photoinduced reversal of the stripe-induced phonon softening observed around the charge-order wave vector by neutron scattering (*18*). For thermal stripe melting in equilibrium, this phonon hardening is likely masked by a redshift from lattice expansion and possibly by changes in the folded phonon energy due to a temperature-dependent stripe wave vector $q_{CO}$.

Probing the transient conductivity around the Ni-O bending mode thus allows us to capture both the electronic and lattice dynamics during the stripe melting process within a single terahertz spectrum. The peak strengths of the zone-folded longitudinal and transverse modes are fingerprints of the presence of corresponding periodic lattice distortions. Hence, the data reported in Fig. 4 (A to C) provide insight into the mutual coupling between the probed subsystems – itinerant electrons along with longitudinal and transverse lattice distortions – whose dynamics can be modeled by a set of differential equations (see the Materials and Methods). The best fit of the dynamics is shown as solid lines in Fig. 4 (A to C). Here, the quench of the longitudinal distortions exhibits a rise time that matches the time constant $\tau_{loc} \approx 0.4$ ps of electronic localization. This reflects a strong coupling of these vibrations to the melting of the electronic charge order. The much slower quench of the transverse distortions can be understood as a vibrational coupling ($\tau_{L-T} \approx 1.5$ ps) between longitudinal and transverse distortions due to anharmonic interactions. A direct coupling of the transverse vibrations to the electronic dynamics instead does not result in a consistent representation of the dynamics (see fig. S9). Finally, the recovery of symmetry breaking is described by a time constant $\tau_{form} \approx 7$ ps and necessitates the dissipation of energy into lower-frequency acoustic modes (*27*) as vibrational stripe order is restored. Both this slow relaxation dynamics and the ≈20% overall signal changes are similar to that of x-ray scattering from electronic stripe order in LSNO at a comparable pump fluence (*19*), indicating a dynamic lock-in of electronic and vibrational order in the final phase of stripe formation.

The dynamics observed here differs strongly from charge and orbital ordering in manganites where the electronic and structural dynamics follow a single time-dependent order parameter (*7*), or in the insulator-to-metal transition in $VO_2$ where vibrational modes promptly disappear after excitation (*6*). Our terahertz probe of stripe melting reveals a picosecond loss of the vibrational superlattice, which also differs from the full decoupling observed in excitonic charge-density waves in dichalcogenides (*28*). Notably, the fast dynamics of the LO mode cannot be explained by transient screening: The photo-induced changes of the conductivity are small and – as shown in Fig. 4D – the bulk LO-TO splitting remains largely unchanged after photoexcitation. Instead, the much different coupling strengths for longitudinal and transverse distortions obtained in our model point to a selective microscopic coupling which is sensitive to the directionality of the lattice motion. This can be understood by considering the strength of polar coupling between electrons and lattice modes: The respective interaction energy is given by the overlap integral $E_{int} \propto \int P_{vib}(r) E_{co}(r) d^3r$ between the optical phonon polarization $P_{vib}$ and the local field $E_{co}$ arising from the stripe's charge density modulation (*29*). Because the LO distortions along $q_{CO}$ are polarized perpendicular to the stripes, they develop a maximal overlap with the periodic charge modulation – thus explaining a strong coupling to the photoinduced electronic quench. In contrast, polar interactions play only a minor role for the TO distortions based on their atomic displacements along the stripes with vanishing charge gradient. The slow dynamics of the transverse distortions instead suggests a sensitivity to the overall lattice relaxation during stripe melting, which can be approximated by the ≈1 ps acoustic propagation time between adjacent stripes. Thermal equilibrium is reached only on much longer times, however, as evidenced by the slow ≈25 ps reaction of the lattice Bragg peak in time-resolved X-ray studies (*20*). Thus, the transient terahertz spectra reveal a multiscale stripe dynamics, which motivates theoretical investigations using advanced quantum models (*30*) for further microscopic insight.

In summary, our investigations of the ultrafast multi-terahertz response around the Ni-O bending mode of the model stripe-ordered system $La_{1.75}Sr_{0.25}NiO_4$ provide novel insight into the time evolution of the electronic conductivity and vibrational symmetry upon melting and formation of atomic-scale charge order. Exploiting sensitive vibrational signatures of periodic lattice distortions, we find that the crystal symmetry remains initially frozen after ultrafast optical excitation of the electronic system. The different reaction times of longitudinal and transverse zone-folded phonons highlight the role of strongly directional polar interactions. The overall delayed vibrational dynamics may affect both light-driven superconductivity and – more generally – the spatiotemporal motion of stripe fluctuations in charge-ordered superconductors with possible implications for the nature of the pairing glue. The multiscale dynamics of electronic and vibrational stripe constituents revealed here thus motivates further ultrafast optical or photoemission studies to investigate transient spectral and momentum-space fingerprints of nanoscale charge and lattice order in correlated oxides.

## MATERIALS AND METHODS

**Samples and equilibrium characterization**

We investigated $La_{1.75}Sr_{0.25}NiO_4$ grown as single crystals via the floating-zone method, and subsequently cut and polished along the (1 0 0) surface. Vectors are given in orthorhombic notation, whereas [1 0 0] and [0 1 0] are rotated 45º with respect to the crystallographic axes (Ni-O bonds) of the tetragonal lattice, while [0 0 1] corresponds to the c-axis. At the doping level studied here, long-range stripe patterns of holes form in the Ni-O plane below the charge-order transition temperature $T_{CO} \approx 105$ K and are accompanied by spin order below ≈90 K, as determined by XRD. The stripes are oriented either along [1 0 0] or [0 1 0], forming a twinned distribution of 90º-rotated nanoscale domains with ≈20 nm correlation length. Therefore, an incident terahertz wave polarized along the [0 1 0] direction is aligned either parallel or perpendicular to the stripe direction in each nanodomain.

The equilibrium reflectivity $R(\omega)$ was measured at different temperatures with a Bruker 66v/S vacuum FTIR spectrometer at the Advanced Light Source (Beamline 1.4.2) polarized along the Ni-O plane under near-normal incidence. The data were taken in a large spectral range from 0.01 – 1 eV, whereas outside this range it was smoothly continued from known optical data of nickelates at similar dopings (*22*) and by model functions that approximate the extreme low- and high-frequency limits. The optical conductivity was then obtained from the broadband reflectivity via the Kramers-Kronig constrained variational analysis (*31*).

**Ultrafast terahertz experiments and data analysis**

Optical-pump terahertz-probe experiments were performed in reflection geometry. For this, the train of 50-fs pulses from a 250-kHz Ti:sapphire regenerative amplifier (Coherent RegA) was split into three beams to enable near-IR photoexcitation along with terahertz generation and detection (*32*). To achieve high sensitivity and spectral resolution, we used phase-matched terahertz generation and electro-optic sampling in 0.5-mm thick GaSe crystals around ≈40º angle of incidence. Both pump and probe beams were polarized along the Ni-O plane. The pump was focused to a ≈200 μm diameter (FWHM, full width at half maximum), whereas the probe diameter was ≈120 μm (FWHM). The incident pump fluence used in this work was ≈ 0.5 mJ/cm$^2$, resulting in a maximum estimated lattice heating of $\Delta T \approx$ 33 K at long delay times (for details see Supplementary section S1).

In the pump-probe experiments, the transient pump-induced change $\Delta E(t)$ of the terahertz field reflected from the LSNO crystal was measured via electro-optic sampling. For each fixed delay $\Delta t = t - t_{pu}$ between the sampling and pump time points, the change $\Delta E(t)$ was measured by varying the terahertz generation delay using the scheme of Kindt and Schmuttenmaer (*24*) to avoid spectral and temporal artifacts (for details see section S2). For each delay $\Delta t$, the transient conductivity change $\Delta\sigma(\omega)$ was then obtained from the terahertz field change $\Delta E(t)$ and reference field $E(t)$, using a transfer matrix method to account for the excitation depth profile and penetration depth mismatch between the IR pump and terahertz probe. The pump penetration depth is $\delta_{pump} \approx$125 nm, whereas that of the terahertz probe is larger and varies across the resonance as $\delta_{probe} \approx$ 0.5 – 5 μm. Thus, we consider an exponentially-decaying change of the complex-valued index of refraction $n_{exc}(\omega, z) = n_{eq}(\omega) + \Delta n(\omega)\exp(-z/\delta_{pump})$, where $z$ is the distance along the sample normal. The complex-valued field reflection coefficient $\phi_r(\omega)$ of the photoexcited sample was calculated by finely approximating the exponentially-decaying changes with a multilayer structure of up to 100 layers, using the transfer matrix formalism (*33*). For each terahertz frequency, we then numerically solved for the complex-valued $\Delta n(\omega)$ that reproduces the measured field change $\Delta E_{THz}(\omega)$. Corresponding changes of the optical conductivity $\Delta\sigma$ and dielectric function $\Delta\varepsilon$ were then derived from $\Delta n$.

**DFT calculations**

Calculations of the electronic structure of the La$_2$NiO$_4$ lattice were performed on the basis of density functional theory with the PBE exchange functional (*34*), using the full potential (linearized) augmented plane-wave method as implemented in the WIEN2k (*35*) package and a *k*-mesh size of 11 x 11 x 7 for the self-consistent calculations. The lattice is based on the undoped I4/mmm structure (*36*). To determine the phonon dispersion, we employed Phonopy to calculate the spectra based on the finite displacement method for a 2x2x1 supercell (*37*). Moreover, the nonanalytic correction due to retarded Coulomb interactions was calculated from a dispersion using VASP and approximated with a Thomas-Fermi potential, resulting in the dispersion in Fig. 1D (dashed line) that accounts for the experimentally observed LO-TO splitting for small phonon wave vectors.

**Multi-oscillator phonon model**

The vibrational spectra were modeled with a multiplet of *N* Lorentz oscillators centered around the high-temperature bending mode frequency, with the complex dielectric function expressed as

$$\varepsilon(\omega) = \varepsilon_b + \sum_{j=1}^{N} \frac{\omega_{p,j}^2}{(\omega^2 - \omega_j^2) - i\Gamma_j \omega}. \tag{1}$$

Here, $\varepsilon_b$ is the background dielectric constant, and $\omega_{p,j}$, $\omega_j$, and $\Gamma_j$ are, respectively, the oscillator's plasma frequency, resonance frequency and linewidth. From the dielectric function we directly obtained the reflectivity and conductivity for comparison with the experimental data.

In the differential multi-oscillator fitting procedure in Fig. 3, the spectral weight $\omega_{p,j}$ and relaxation rate $\Gamma_j$ both influenced the peak strength $\omega_{p,j}^2/\Gamma_j$. Because it was not possible to distinguish these quantities independently with the present energy resolution, we constrained their relative variations to be identical in magnitude. Different constraints do not alter significantly the $\chi^2$ and outcome of the fitting procedure.

**Dynamic coupling model**

The dynamics of the itinerant carriers along with the longitudinal and transverse distortions can be approximated with a set of differential equations describing their mutual coupling:

$$\frac{d}{dt}\Delta\eta_{\text{el}}(t) = I_{\text{pump}}(t) - \tau_{\text{loc}}^{-1}\Delta\eta_{\text{el}}(t) \tag{2}$$

$$\frac{d}{dt}\Delta\eta_{\text{LO}}(t) = \beta \cdot \Delta\eta_{\text{el}}(t) - \tau_{\text{L-T}}^{-1}[\Delta\eta_{\text{LO}}(t) - \Delta\eta_{\text{TO}}(t)] - \tau_{\text{form}}^{-1}\Delta\eta_{\text{LO}}(t) \tag{3}$$

$$\frac{d}{dt}\Delta\eta_{\text{TO}}(t) = \tau_{\text{L-T}}^{-1}[\Delta\eta_{\text{LO}}(t) - \Delta\eta_{\text{TO}}(t)] - \tau_{\text{form}}^{-1}\Delta\eta_{\text{TO}}(t) \tag{4}$$

Here, $\Delta\eta_{el}$ is the relative perturbation of the electronic conductivity, whereas $\Delta\eta_{LO}$ and $\Delta\eta_{TO}$ represent the relative perturbations of the zone-folded LO and TO peak strengths as indicators of the corresponding stripe-symmetry lattice distortions. Moreover, $I_{pump}(t)$ represents the near-IR photoexcitation of itinerant carriers, whereas their decay is given by the charge localization time $\tau_{loc}$. As represented in Eq. (3) these decaying carriers $\Delta\eta_{el}(t)$ act as a source term for melting the order, resulting in a matched quench time of the LO distortions (with the amplitude scaled by $\beta < 0$). Electronic coupling to the TO distortions yields an inferior description and was thus omitted (compare to fig. S9). The TO distortions were instead coupled to the longitudinal perturbations with a time constant $\tau_{L-T}$, representing an anharmonic vibrational coupling. Finally, the recovery of stripe-phase vibrational symmetry was reproduced with a common time constant $\tau_{form}$ in the model.

## SUPPLEMENTARY MATERIALS

Supplementary material for this article is available at:
http://advances.sciencemag.org/cgi/content/full/3/11/e1600735/DC1

section S1. Thermal heating and excitation transport.
section S2. Time and frequency resolution in the optical-pump THz-probe experiments.
Fig. S1. Optical-pump terahertz-probe scheme.
Fig. S2. Energy level scheme and couplings used in the simulation.
Fig. S3. Simulated transient terahertz spectra resulting from different probe schemes.
Fig. S4. Equilibrium optical reflectivity around the bending mode of $La_{1.75}Sr_{0.25}NiO_4$.
Fig. S5. Temperature dependence of the sharp modes below $T_{CO}$.
Fig. S6. Fluence dependence of the mid-IR pseudogap dynamics.
Fig. S7. Dynamics of the $\omega_3$ zone-folded vibrational mode.
Fig. S8. Dynamics of the $q = 0$ bending phonon mode.
Fig. S9. Model comparison for the TO mode dynamics.
References *(38–46)*

**Acknowledgements:** We thank W.-S. Lee and Z.-X. Shen for discussions and contributions in the early stage of this work, as well as A. Cavalleri and M. Trigo for stimulating scientific discussions.

**Funding:** This work was primarily funded by the U.S. Department of Energy, Office of Science, Office of Basic Energy Sciences, Materials Sciences and Engineering Division under contract no. DE-AC02-05CH11231 (Ultrafast Materials Science program KC2203), covering ultrafast and equilibrium terahertz spectroscopy and data analysis (G.C., S.B., B.H., A.L., and R.A.K.). For equilibrium FTIR characterization, the research used resources of the Advanced Light Source, which is a DOE Office of Science User Facility supported under Contract No. DE-AC02-05CH11231 (H.A.B. and M.C.M.). B.H. and S.B. acknowledge student fellowships from the German Academic Exchange Service.

**Author Contributions:** G.C. and R.A.K. conceived and performed the ultrafast terahertz study; A.F.K. performed the DFT calculations; G.C., B.H., H.A.B. and M.C.M. implemented the FTIR spectroscopy; T.S. synthesized the crystals; and G.C. and R.A.K. analyzed the data and wrote the manuscript. All the authors contributed to the discussion and interpretation of the experimental data.

**Competing interests:** The authors declare that they have no competing interests.

**Data and materials availability:** All data needed to evaluate the conclusions in the paper are present in the paper and/or the Supplementary Materials. Additional data related to this paper may be requested from the authors.

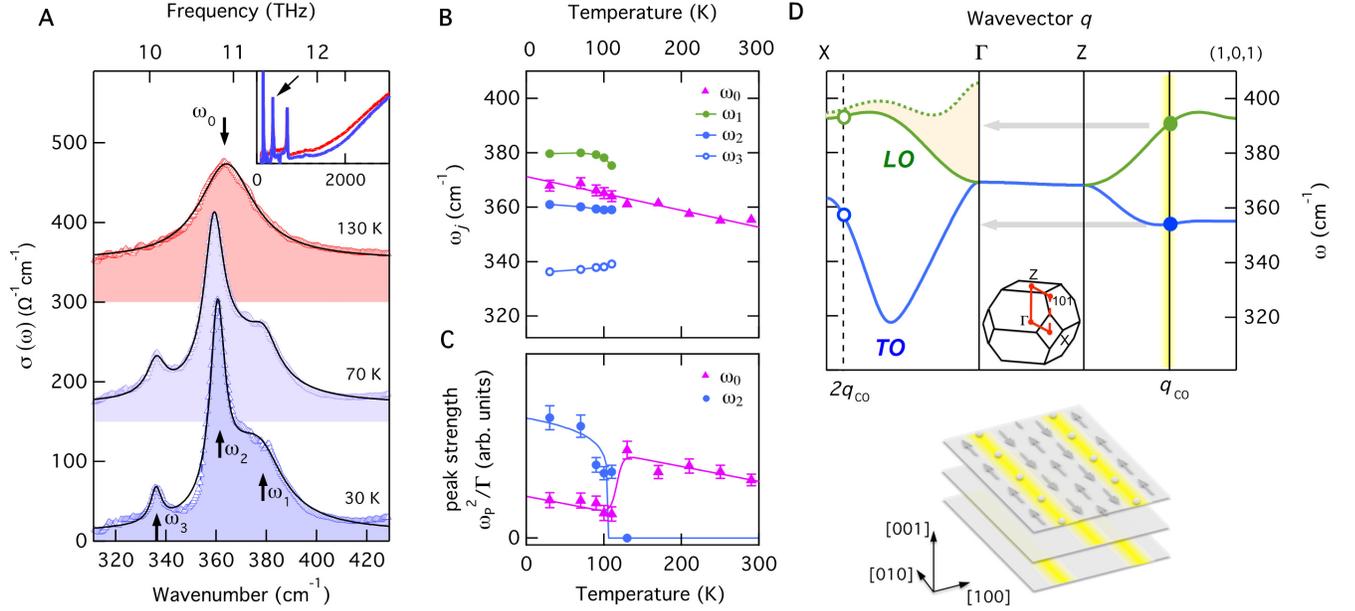

**Fig 1. Equilibrium optical conductivity around the Ni-O bending mode of La$_{1.75}$Sr$_{0.25}$NiO$_4$.** (**A**) In-plane optical conductivity $\sigma_1(\omega)$ of LSNO at different temperatures (dots), and best fits (lines) with the multi-oscillator model described in the text and the Materials and Methods. Curves are offset vertically for clarity, and the inset shows an expanded range conductivity with additional data in fig. S4. (**B** and **C**) Temperature dependence of the frequency and peak strength from the multi-oscillator fits. Lines are guide to eyes. Error bars are 1 SD. (**D**) Dispersion of the Ni-O bending mode as obtained from density functional theory (DFT) calculations (see the Materials and Methods). Charge ordering reduces the size of the Brillouin zone, which corresponds to the zone-folding of both the transverse optical (TO) and longitudinal optical (LO) modes from $q_{CO} = (0.55,0,1)$ back to $\Gamma$, with coordinates given in an orthorhombic notation (see Materials and Methods). Solid dots indicate the zone-folded modes at $q_{CO}$. Open dots mark second-order modes corresponding to an in-plane momentum $2q_{CO} = (1.1,0,0)$, with some uncertainty on the TO energy due to the strong dispersion. Dotted line indicates LO dispersion including long-range Coulomb interactions, accounting for the observed bulk LO-TO splitting in Fig. 4D. The charge and spin stripe arrangements of LSNO are shown schematically.

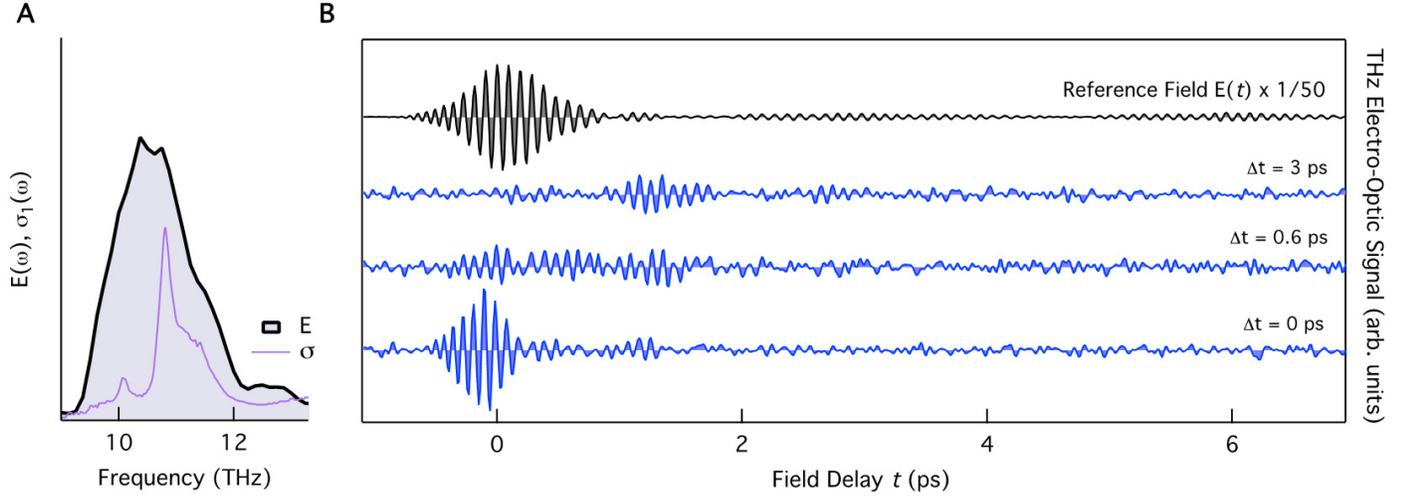

**Fig. 2. Photoinduced terahertz response. (A)** Spectrum of the terahertz probe along with the optical conductivity of the Ni-O bending mode at 30 K. **(B)** Reference terahertz field $E(t)$ (blue) reflected off the unexcited crystal, along with photoinduced field changes $\Delta E(t)$ at different pump-probe delays $\Delta t$. The LSNO crystal was excited at 800 nm wavelength with ≈0.5 mJ/cm$^2$ incident pump fluence corresponding to 15 J/cm$^3$ absorbed energy density (see section S1). The electro-optically sampled fields are plotted as a function of the field delay time $t$, with curves offset vertically for clarity. The oscillations at long delay times – ringing structures - indicate sharp features in the reflected fields. The reference field presents such oscillations only at low temperatures where sharp resonances appear in the reflectivity spectrum (see fig. S4).

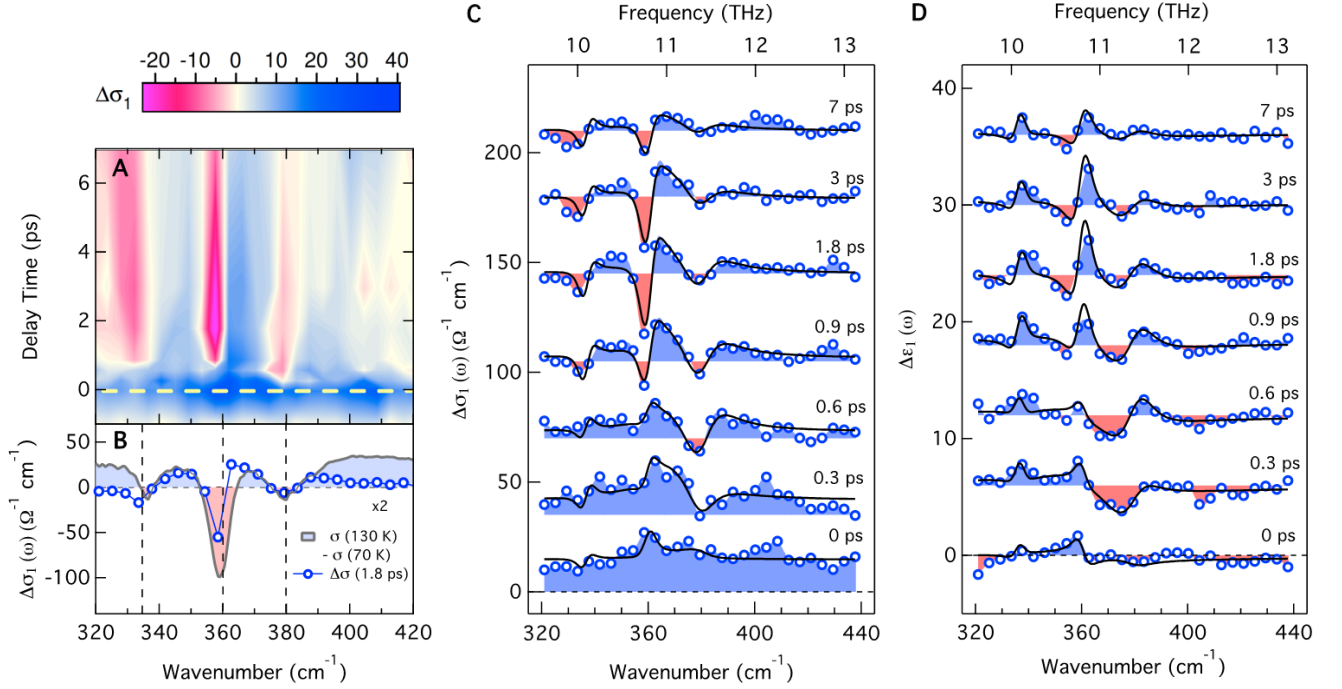

**Fig. 3. Transient optical conductivity changes and model fits.** **(A)** Two-dimensional plot of the photoinduced change of the real part of optical conductivity, $\Delta\sigma_1(\omega)$, in the stripe phase of LSNO as a function of pump-probe delay time $\Delta t$ and probe frequency. **(B)** Photoinduced conductivity change at 1.8 ps (dots) along with the thermal difference from below to above $T_{CO}$ (lines). **(C** and **D)** Transient changes of the real parts of the optical conductivity and dielectric function at several pump-probe delays (dots). The black lines represent the best fit from the differential multi-oscillator model, with corresponding parameters shown in Fig. 4 (A to C) and figs. S7 and S8.

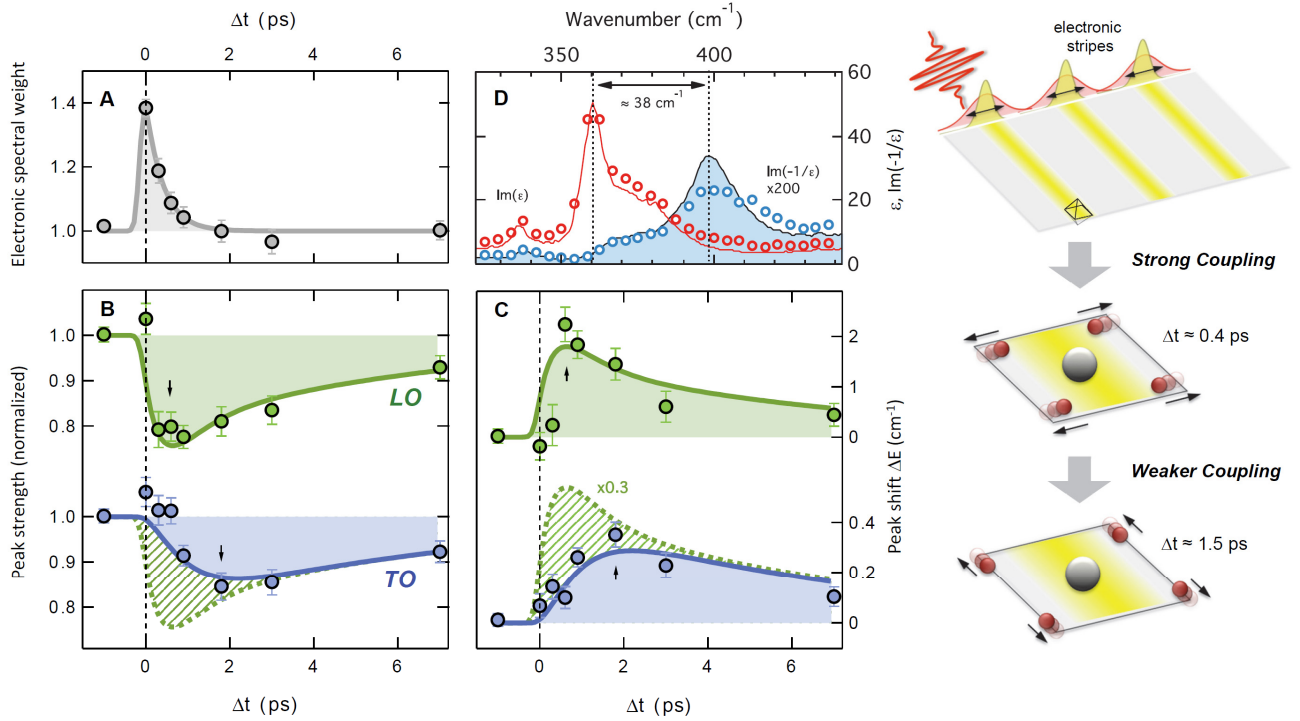

**Fig. 4. Electronic conductivity and vibrational symmetry breaking dynamics.** Dynamics of the multi-oscillator model parameters: **(A)** electronic spectral weight, **(B)** peak strengths of the zone-folded phonon resonances, along with **(C)** their energy shifts. Error bars are 1 SD. Lines indicate the best fit of the data in (A) and (B) with a dynamic coupling model (see the Materials and Methods). The same model (rescaled) is shown for comparison also in (C) where it reproduces the transient energy shifts. **(D)** Imaginary part of the dielectric function Im($\varepsilon$) and energy-loss function Im($-1/\varepsilon$) in equilibrium at 30 K (lines) and upon excitation at $\Delta t = 0$ (open circles), with the bulk LO-TO splitting indicated (arrow). The multiscale dynamics is illustrated on the right-hand side: LO lattice distortions react quickly to the charge order melting due to strong coupling via polar Coulomb interactions, whereas electric fields vanish along the stripes resulting in a delayed reaction of the TO distortions.